\documentclass{resonance}

\usepackage{amsfonts}
\usepackage{amsmath}
\usepackage{amssymb}
\newcommand{\bed}{\begin{displaymath}}
\newcommand{\eed}{\end{displaymath}}
\newcommand{\bei}{\begin{itemize}}
\newcommand{\eei}{\end{itemize}}
\newcommand{\bef}{\begin{figure}}
\newcommand{\eef}{\end{figure}}
\newcommand{\ben}{\begin{enumerate}}
\newcommand{\een}{\end{enumerate}}
\newcommand{\beq}{\begin{equation}}
\newcommand{\eeq}{\end{equation}}
\newcommand{\ber}{\begin{eqnarray}}
\newcommand{\eer}{\end{eqnarray}}

\newcommand{\gcc}{\mbox{${\rm g} \, {\rm cm}^{-3}$}}

\newcommand{\gsim}{\raisebox{-0.3ex}{\mbox{$\stackrel{>}{_\sim} \,$}}}

\newcounter{attnctr} 

\begin{document}


\title{Gravity Defied \\
     From potato asteroids to magnetised neutron stars}
\secondTitle{I. The self-gravitating objects}
\author{Sushan Konar}

\maketitle
\authorIntro{\includegraphics[width=2cm,angle=90]{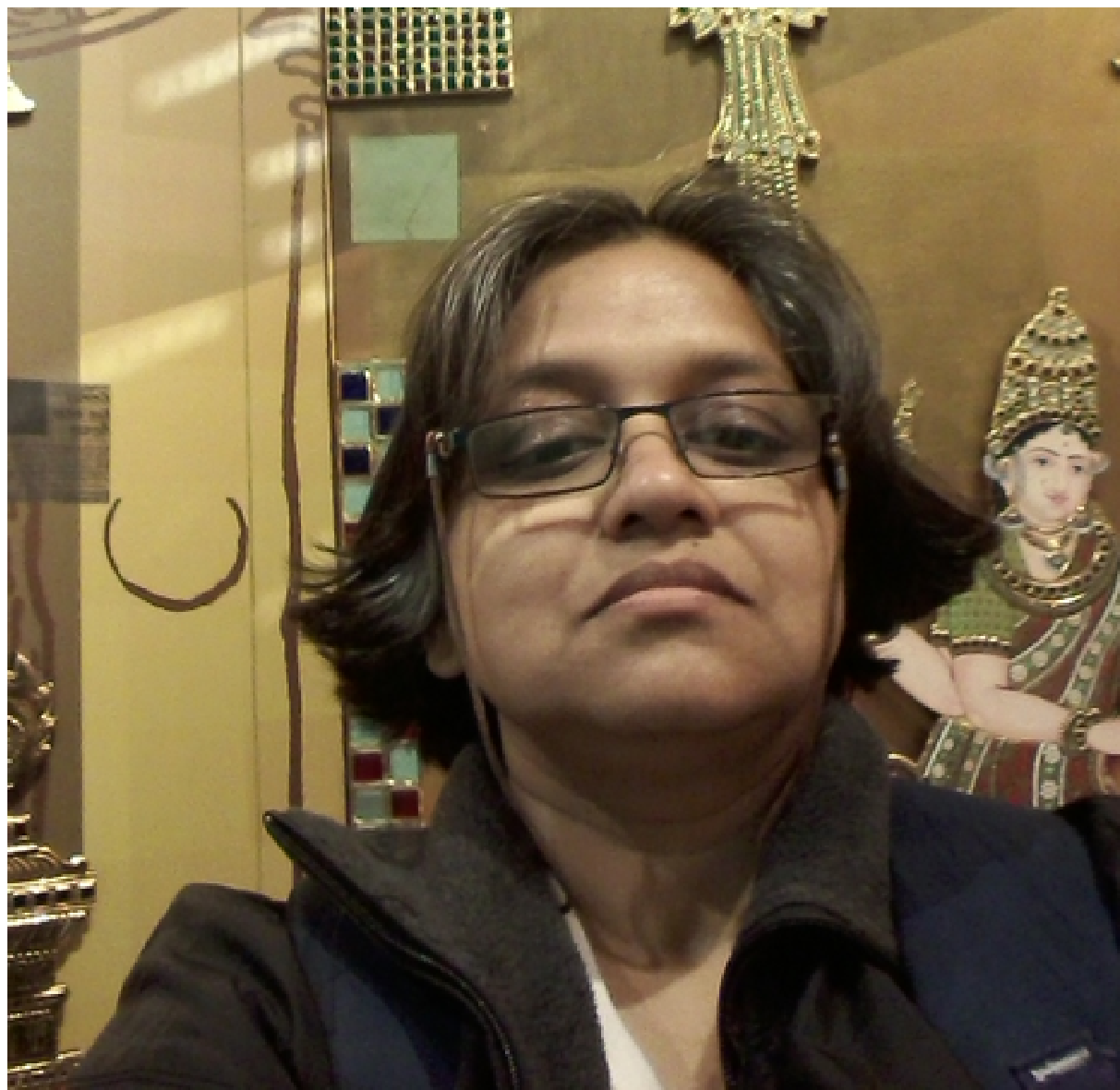}\\
Sushan Konar  works at  NCRA-TIFR, Pune. She  tries to  understand the
physics of stellar  compact objects (white dwarfs,  neutron stars) for
her livelihood and writes a blog about life in academia as a hobby.}
\begin{abstract}
  Gravitation, the  universal attractive  force, acts upon  all matter
  (and radiation)  relentlessly.  Left  to itself, gravity  would pull
  everything  together  and  the  Universe  would  be  nothing  but  a
  gigantic black  hole. Nature throws  almost every bit of  physics -
  rotation,  magnetic  field, heat,  quantum  effects  and so  on,  at
  gravity to escape  such a fate. In this series  of articles we shall
  explore systems where the eternal pull  of gravity has been held off
  by one or another such means.
\end{abstract}
\monthyear{January 2017}
\artNature{GENERAL  ARTICLE}

\section*{Introduction}
It is well  known that each and every popular  lecture on astrophysics
invite  questions on  black holes,  whatever the  actual topic  of the
lecture may be.   This abiding interest is simply  because black holes
are  the most  exotic  of  all astrophysical  objects,  the so  called
`unseen' last frontiers.  Though we have, at last, been able to `hear'
them - as a pair of black  holes merged to form another bigger one.  I
am, of course, referring to  the first ever detection of gravitational
waves, generated in  two such events creating  an unprecedented splash
in the entire scientific world last year.


A black  hole is  a singularity  in space-time  where matter  has been
totally defeated by gravity, whereas in everything else matter retains
its identity even when gravity dominates (perhaps squeezed into exotic
phases,  as expected  in white  dwarfs and  neutron stars).   However,
gravity starts  out its journey  by being  the weakest of  all forces,
gaining  importance only  with  increasing mass.   In  this series  of
articles  we shall  explore  this journey,  looking  at objects  where
gravity is either  insignificant or can be resisted  by another force,
and does  not win over matter  (as it does in  black holes).  Starting
from  tiny atoms  we shall  consider sub-stellar  and stellar  objects
where  material  property  prevails,   before  such  information  gets
completely lost in the interior of a black hole.

There  exist  four  fundamental  forces  in  the  Universe  -  strong,
electromagnetic,  weak  and  gravitational;  in  the  order  of  their
strength of  interaction. The  strong force, more  than two  orders of
magnitude  stronger than  the electromagnetic,  is effective  within a
distance  scale of  $\sim 10^{-13}$~cm  and falls  off rapidly  beyond
this.   Therefore   it  has  virtually  no   presence  beyond  nuclear
dimensions.  The weak force has an even shorter range - expected to be
$\sim 1\%$  of the proton radius.  Moreover, both of these  forces are
effective for only  a certain kind of particles.  That  leaves us with
the  electromagnetic  and the  gravitational  forces  - ones  that  we
usually encounter in our everyday life.

Interestingly, both of these forces are given by the  generic form,
\beq
   F = K_F \, \frac{q_1 q_2}{r^2} \,,
\eeq
where  $K_F$  is  a  force-specific   constant,  $q_1,  q_2$  are  the
corresponding   charges  and   $r$   is  the   distance  between   the
particles. The forces are directed along $r$. In cgs units these force
equations reduce to,
\ber
   F_{\rm EM} &=& \frac{e_1 e_2}{r^2} \; \; \; \; (\mbox{electromagnetic})\,, \\
   F_{\rm G}  &=& G \, \frac{m_1 m_2}{r^2} \; \; \; \; (\mbox{gravitational}) \,; 
\eer
where  $e_1,  e_2,  m_1,  m_2$  are the  charges  and  masses  of  the
interacting  particles   and  $G$   is  the   universal  gravitational
constant. Though  omnipresent, the gravitational  force is so  weak at
smaller scales that it can be  ignored for all practical purposes. For
example, the ratio between these two forces for a pair of protons is,
\ber
\frac{F_{\rm G}}{F_{\rm EM}}
  = \frac{G m_p^2}{e_p^2} \simeq 10^{-39}\,;
\eer
making  the   resultant  interaction   repulsive  as   the  attractive
gravitational force is too tiny to make any difference.

The interesting point to note is  that there exist electric charges of
two different  signs whereas there  is only one kind  of gravitational
charge (i.e,  mass).  This  immediately explains  why gravity  wins at
large scales.  Any imbalance of  charge gives rise to electric fields,
causing movement of charges that ultimately neutralises the imbalance.
As a result, Universe is charge neutral  as a whole and there exist no
large scale electric fields. On the other hand, gravity increases with
increasing mass.  However,  other forces give gravity  a serious fight
at smaller  length (mass) scales  allowing structures to form.  In the
following  sections we  shall consider  these structures  and see  how
gravity progressively takes precedence as we move to larger and larger
scales.

\section{Bound Systems}

When  a  bound  system  is  formed,  as  a  result  of  an  attractive
interaction, it has a lower energy than the sum of the energies of its
unbound constituents. This difference in  energy, released at the time
of formation of the bound system, is known as the {\em binding energy}
($E_{\rm B}$).  There exist different  types of $E_{\rm B}$, operating
over  different  length  and  energy  scales and characteristic  of  the
underlying attractive interaction - the  smaller the size of the bound
system  (for an  identical set  of constituent  particles) the  higher
being its associated $E_{\rm B}$.

\subsection{From nuclei to asteroids}

At  the smallest  scale, strong  force  binds quarks  together into  a
nucleon (neutron or  proton) with $E_B$ in excess  of 900~MeV ($E_{\rm
  B}^{\rm   proton}$  =   928.9~MeV,  $E_{\rm   B}^{\rm  neutron}$   =
927.7~MeV)~\mfnote{1~eV = 1.6  $\times 10^{-12}$~erg}; whereas nuclear
binding  energy  derives  from  residual strong  force,  ranging  from
2.2~MeV per nucleon in deuterium to 8.8~MeV per nucleon for Ni$^{62}$.
Further on, at the atomic  level, the electron binding energy, arising
from  the electromagnetic  interaction between  the electrons  and the
nucleus, is a measure of the  energy required to free an electron from
its orbit and is commonly known  as the ionisation energy (13.6~eV for
atomic Hydrogen).   For an excellent summary  of particle interactions
and  constituents  of  bound   systems  please  see  the  illustration
at~\cite{pdg}.

\begin{figure}[!t]
  \caption{van  der Waals  potential, $V(r)$, combination  of a)  a
    short-range, hard-sphere repulsion and  b) a long range attraction
    as a  function of the  inter-particle distance $r$.  The resultant
    reaches its  attractive minimum  ($- V_0$) at  a distance  of $r_0$
    (effective distance between two hard-sphere particles touching each
    other).}
  \label{f_vdw}
\vspace{-0.5cm} 
\centering\includegraphics[width=6.5cm]{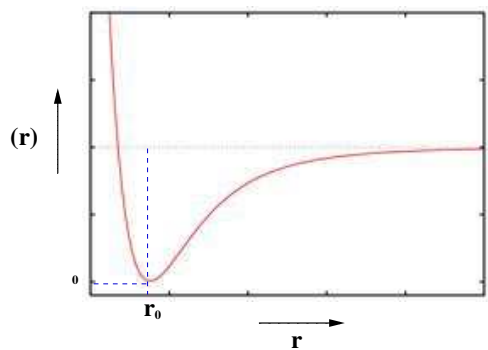}
\vspace{-0.5cm} 
\end{figure}

The attraction  underlying the bond  energy of molecules,  measuring a
few eVs,  is again electromagnetic  though modified by  the electronic
structure of a given molecule.  At low temperatures, the loosely bound
molecules of  a gas condense  into a liquid/solid phase.   The binding
energy,  known as  the cohesive  energy,  is gained  by arranging  the
molecules   into  a   condensed  phase.    It  is   weaker  than   the
intra-molecular  bonds, arising  out  of an  effective  van der  Waals
attraction between neutral particles (Fig.~\ref{f_vdw}) and provides a
measure for the rigidity, since  the energy required for a substantial
deformation  of   any  material  must   be  similar  to   its  binding
energy. Therefore, cohesive energy is responsible for holding ordinary
solids  together  - from  small  chalk  pieces  in our  classrooms  to
odd-shaped asteroids hurtling through space.

\subsection{Beyond the potato asteroids}

\begin{figure}[!t]
\caption{Asteroid shape depends on  size -  tiny  Eros is  completely
  misshapen, while much larger Ceres is almost completely spherical.
  Pictures taken from {\tt http://solarsystem.nasa.gov/}.}
\label{f_ast}
\vspace{-0.5cm} 
\centering\includegraphics[width=10.0cm]{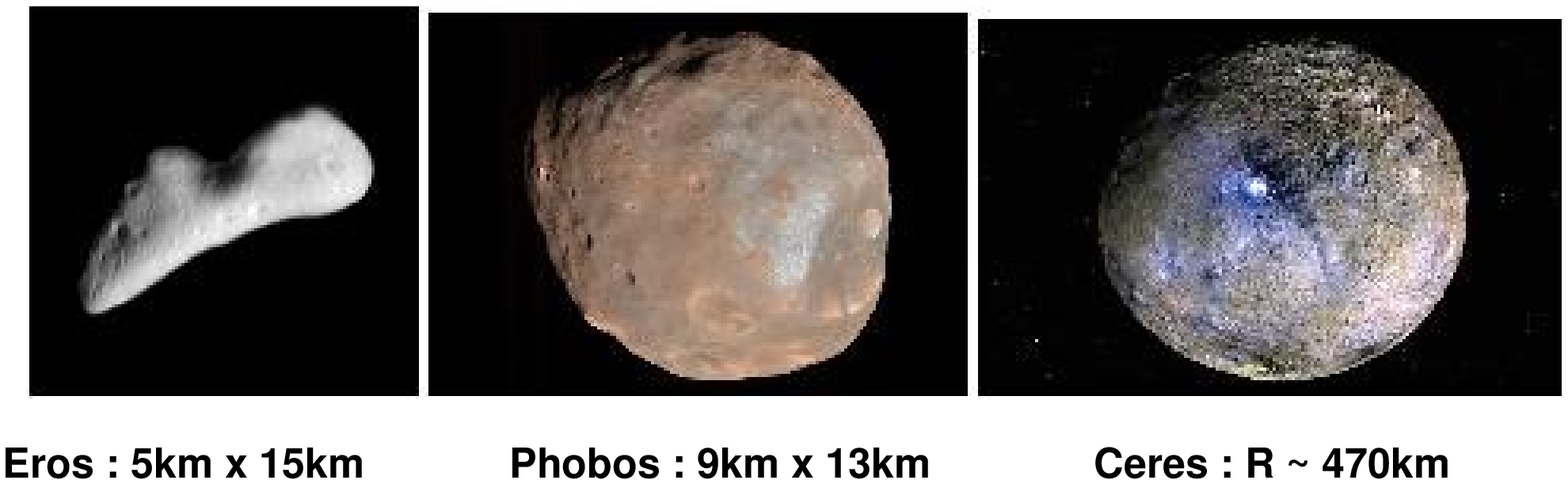}
\vspace{-0.5cm} 
\end{figure}

Sometimes  art  unintentionally  catches   up  with  real  life.   The
designers of  `Stars Wars -  Episode V', that Hollywood  cult classic,
actually used some potatoes for  the asteroid field scene. 
Recent asteroid  exploring missions  have observed that  while smaller
asteroids have irregular `potato' shapes the larger objects are nearly
spherical  - the  transition  happening at  an  approximate radius  of
200-300~km (Fig.~\ref{f_ast}).. This is  known as the {\em potato radius}  ($R_{\rm p}$) -
separating bona fide asteroids from their more spherical counterparts,
the dwarf planets.

\begin{figure}[!t]
  \caption{Surface mountains on a rocky planet. While natural mountains
    (cliff-faced or gently sloping) are bound by the physics discussed
  in the text, artificial step pyramids may not conform.}
  \label{f_mnt}
\vspace{-0.5cm} 
\centering\includegraphics[width=9.0cm]{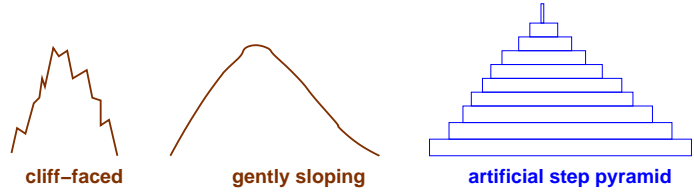}
\vspace{-0.5cm} 
\end{figure}

$R_{\rm p}$ can be obtained from  the elastic property of the asteroid
material.   This is,  in  fact, related  to the  maximum  height of  a
mountain on  the surface of a  solid planet.  For an  arbitrarily tall
mountain, the  gravitational pressure at  the base would  overcome the
yield strength  of the material  and deform  the mountain back  to the
maximum  allowable   height  (Fig.~\ref{f_mnt}).  The   condition  for
stability of such a mountain (assuming average density of the mountain
to be  same as that  at the  base) is that  the pressure at  the base,
given by
\beq
P_{\rm m} = \rho_{\rm m} \, g_{\rm s} \, h_{\rm m},
\eeq
is  less than  the  shear stress  of the  material  ($\rho_{\rm m}$  -
average mountain density, $g_{\rm s}$ - surface gravity of the planet,
$h_{\rm  m}$ -  mountain height).   Hence, the  height of  the tallest
mountain is,
\beq
h_{\rm m}^{\rm max} = \frac{\sigma}{\rho_{\rm m} \, g_{\rm s}},
\label{eq-hm}
\eeq
where $\sigma$ is the shear stress of the planetary material. In terms
of the planetary radius ($R_{\rm Pl}$) this reduces to,
\beq
h_{\rm m}^{\rm max} = \frac{3 \, \sigma}{4 \pi \, G \, \rho_{\rm Pl} \, R_{\rm Pl}} \,;
\eeq
assuming  $\rho_m$ to  be equal  to the  average planetary  density ($
\rho_m  \simeq \rho_{\rm  Pl} =  3 M_p/4  \pi R_{\rm  Pl}^3$) and  the
surface gravity ($g_{\rm s} = G \, M_{\rm Pl}/R_{\rm Pl}^2$) to remain
constant over the  height of the mountain (M$_{\rm Pl}$  - mass of the
planet). For  Earth, this  formula (using appropriate  average density
and shear stress) gives a value  remarkably close to the height of the
Mt. Everest!

An oblong asteroid can be thought  of as a small spherical planet plus
a  large  surface  mountain. According  to  Eq.[\ref{eq-hm}],  $h_{\rm
  m}^{\rm max}$  would decrease with  an increase in $R_{\rm  Pl}$ and
would eventually become smaller than  $R_{\rm Pl}$, at which point the
asteroid should become approximately  spherical. Therefore, the potato
radius,  $R_{\rm p}$,  is attained  when the  maximum mountain  height
equals the radius of the asteroid and is given by,
\beq
R_{\rm p} = \sqrt{\frac{3 \sigma}{4 \pi G \rho_{\rm Pl}}} \,.
\label{eq-rp}
\eeq
Evidently, objects larger than $R_{\rm p}$ are nearly spherical, while
smaller objects can have non-spherical shapes because they do not have
sufficient  gravity  to overcome  their  intrinsic  rigidity. This  is
indicative of the fact that beyond $R_{\rm p}$, gravity dominates over
cohesive energy in what are known as the `self-gravitating' objects.

\section{Self-gravitating Objects}

A `self-gravitating' object is defined to  be bound by its own gravity
- the binding energy coming from  the gravitational interaction of its
constituents.  Earlier,  we  have  seen that  gravity  dominates  over
cohesive  energy  in  objects  with  dimensions  larger  than  $R_{\rm
  p}$. These larger objects would  then be `self-gravitating'.  Let us
see if we arrive at the same conclusion from both of these directions.

The gravitational energy,  $E_{\rm G}$, of a spherical  object of mass
$M$ and radius $R$ is approximately given by,
\beq
E_{\rm G} \simeq \frac{3 G M^2}{5 R} \,.
\eeq
Let us assume that the cohesive  energy of the constituent material is
$\epsilon_{\rm c}$  per unit mass.  Then the total cohesive  energy of
the object is
\beq
E_{\rm c} \simeq  \epsilon_{\rm c} M \,.
\eeq
The condition for self-gravitation would then be given by,
\beq
E_{\rm G} \, \gsim \, E_{\rm c} \, \Rightarrow \,
R \, \gsim  \, \sqrt{\frac{3}{4 \pi G \rho} \epsilon_{\rm c}} \,.
\eeq
This is  exactly equal  to $R_{\rm p}$  derived in  Eq.[\ref{eq-rp}] -
because the cohesive energy, $\epsilon_{\rm  c}$, is of the same order
of magnitude  as the  shear stress,  $\sigma$, of  a given  solid.  In
other  words,  objects  with  radius   larger  than $R_{\rm  p}$  are
self-gravitating.  Assuming  all  the  asteroids,  dwarf  planets  and
terrestrial planets (we  shall talk about Jovian planets  later) to be
composed of similar rocky material ($\rho \sim 5$~\gcc, $\epsilon_{\rm
  c} \sim 10^9$~cgs  units), $R_{\rm p}$ turns out to  be in the range
of  200-300~Km, exactly the radius at which smallest  dwarf planets
with spherical shapes have been observed. 

Interestingly,  gravitaionally bound  systems have  a rather  peculiar
property  -  they  have  negative specific  heat.   Consider  a  small
particle of mass  $m$, going around a  larger mass $M$ (at  rest) in a
circular orbit of radius $d$.  Then the balance of force dictates,
\beq
\frac{G M m}{d^2} = \frac{m V^2}{d}
\label{eq-force}
\eeq
where $V$ is the linear velocity of the mass $m$. The total energy of the
system, given by the sum of its kinetic and potential energy, is
\beq
E = \frac{1}{2} m V^2 - \frac{G Mm}{d} = - \frac{1}{2} \frac{G Mm}{d} \,.
\eeq
This is  true of every  orbit at any  arbitrary distance $d$  from the
central mass $M$.  Suppose now the velocity of the smaller mass $m$ is
increased (by  an wish-granting genie,  perhaps) to $V'$, where  $V' >
V$.  As a result  of this the orbital radius would  change to $d'$,
where $d' < d$ according to Eq.[\ref{eq-force}] and the total energy
would change to
\beq
E' = - \frac{1}{2} \frac{G Mm}{d'} = - \frac{1}{2} m V'^2 < E \,.
\eeq
Therefore,  an increase  in the  velocity has  resulted in  an overall
decrease in the  total energy (and a shrinking of  the orbit).  If the
velocity  of  an object  is  considered  to  be  an indicator  of  its
temperature (the  rms velocity  of the  particles of  an ideal  gas is
proportional  to the  temperature  of  the gas)  then  an increase  in
(effective) temperature  of the system  has resulted in a  decrease in
the   internal   energy.    Thermodynamically  speaking   then   -   a
gravitationally bound system has a  negative specific heat. (Later, we
shall  see that  the  temperature of  a  gravitationally bound  object
actually rises as a result of contraction.)

\subsection{Terrestrial vs. Jovian planets}

As far as  solid objects are concerned we have  already separated them
into two classes - with dimensions  smaller and larger than the potato
radius.   The  larger objects  (all  the  planets, dwarf  planets  and
satellites) are  self-gravitating. However,  even among  planets there
appears  to exist  another  classification -  terrestrial and  Jovian.
Loosely speaking, rocky objects with thin or no atmospheres are termed
`terrestrial'  (Earth-like), while  large gaseous  objects with  thick
atmospheric  covers  and  icy  interiors are  known  as  `Jovian'
(Jupiter-like)  planets.    Interestingly,  the  composition   of  the
atmospheres (wherever  they exist)  also appear  to be  very different
(see  table[\ref{tble-atms}])  in  these  two classes.  No  marks  for
guessing  that this  is nothing  but  a manifestation  of yet  another
tug-of-war between gravity and some other physical force.

\begin{table}[!t]
  \caption{Escape velocity ($V_{\rm E}$), average surface temperature
    ($T_{\rm s}$) and atmospheric compositions of planets in the Solar system.}
\label{tble-atms}
\vspace{-0.25cm}
\centering
\begin{tabular}{llcrl}
          & planet & $V_{\rm _E}$ (km/s) & $T_{\rm s}$ ($^\circ$C) & atmosphere \\
\underline{\em Terrestrial}
& Mercury & 4  & 260  & \\
& Venus   & 10 & 480  & CO$_2$\\
& Earth   & 11 & 15   & N$_2$, CO$_2$\\
& Mars    & 5  & -60  & CO$_2$\\
\underline{\em Jovian} 
& Jupiter & 60 & -150 & H$_2$, He\\
& Saturn  & 36 & -170 & H$_2$, He\\
& Uranus  & 21 & -200 & H$_2$, CH$_4$\\
& Neptune & 23 & -210 & H$_2$, CH$_4$\\
\underline{\em Dwarf}
& Pluto   & 1  & -220 & CH$_4$\\
\end{tabular}
\end{table}

According to the  standard theory of solar system  formation, both the
Sun and  the planets formed from  a primordial gas cloud.  The primary
atmospheres of the terrestrial planets as well as those in the Sun and
the  Jovian  planets were  quite  similar.   The composition  of  this
atmosphere is guessed to be $\sim 94\%$ of atomic hydrogen, $\sim 6\%$
of  atomic helium  and  $\sim  0.1\%$ of  other  gases.  However,  the
terrestrial  planets  mostly lost  this  primary  atmosphere. This  is
related to the escape velocity of the planets.

The  escape velocity,  $V_{\rm E}$,  is the  smallest velocity  that a
particle must  have to escape  from the gravitational attraction  of a
massive  object.  It  means that  the kinetic  energy of  the particle
should be more than its potential energy in the gravitational field of
the massive object.   Consider a small mass $m$,  in the gravitational
field of a  large mass $M$ of  radius $R$. For $m$ to  escape from the
surface of $M$, it should have a velocity $V$ such that,
\beq
\frac{1}{2} m V^2 \geq \frac{GMm}{R}  \Rightarrow V_{\rm E} = \sqrt{{2
    GM}{R}} \,.
\eeq
Clearly,  larger escape  velocities are  required to  escape from  the
gravitational field  of larger masses (see  table[\ref{tble-atms}] for
escape velocities from solar system planets).

Evidently,  an atmospheric  particle  would not  remain  bound to  the
planet if it  happens to have an average velocity  that is larger than
or equal to the  escape velocity of that planet. We  know that the rms
velocity,  $V_{\rm  rms}$,  of  a   gas  particle  of  mass  $m_g$  is
proportional to the temperature, $T$, of that gas and is given by
\beq m_g V_{\rm  rms}^2 \propto k_{\rm B} T \Rightarrow  V_{\rm rms} \propto
\sqrt{\frac{k_B T}{m}}\,,
\eeq
where $k_{\rm B}$  is the Boltzmann constant.  It is  to be noted that
the lighter  particles would  have higher $V_{\rm  rms}$ for  the same
temperature.  The  temperature of a planetary  atmosphere is basically
determined by the surface temperature of that planet (the layer of gas
in contact  with the surface would  have the same temperature  and the
temperature would drop exponentially  with distance from the surface).
Therefore,  for large  enough  surface temperatures  and small  enough
escape velocities (for  $V_{\rm rms} \geq V_{\rm  E}$) the atmospheric
particles would escape the gravitational pull of a particular planet.

The  inner planets,  being  closer  to the  Sun,  have higher  surface
temperatures whereas the outer planets are cooler. Moreover, the inner
planets  are  lighter  compared   to  their  Jovian  counterparts  and
therefore  have smaller  escape  velocities.  As  a  result the  outer
Jovian planets retain their primary atmosphere, whereas in case of the
inner terrestrial  planets most  of the  original hydrogen  and helium
have been lost. In addition, the  abundance of hydrogen in cold Jovian
planets  has  allowed  it  to  combine with  other  elements  to  form
compounds  like  methane and  ammonia.  All  these factors  have  been
responsible  in  modifying  the  atmospheres -  resulting  in  totally
different compositions in terrestrial and Jovian planets.

\section{Going forward..}

Obviously  Jovian   planets  are  quite  different   from  their  rocky
terrestrial  counterparts. Instead,  they  are gaseous  and more  like
colder versions of the  Sun (composition-wise) itself.  The similarity
actually  goes deeper  and  I prefer  of think  of  these objects  (in
particular, Jupiter and Saturn) as `could have been stars' rather than
planets.  Of course,  that is a story  for another day, to  be told in
the next article of this series.

\section*{Acknowledgment}

My thanks to Rajaram Nityananda  for bringing Mt. Everest right inside
our  astrophysics class  and to  Varun Bhalerao  for telling  me about
`potato' asteroids; also, to the anonymous referee for making a number
of valuable suggestions  and to the editors of  this `special'
issue of Resonance for making a `special' effort.


\begin{thebibliography}{99} 
\bibitem{potato} 
  M. E. Caplan,
  \textit{Calculating the Potato Radius of Asteroids using the Height of Mt. Everest},
  arXiv:1511.04297, 2015
\bibitem{pdg} 
{\textit http://www.physi.uni-heidelberg.de/Einrichtungen/FP/anleitungen/F13/gif/cpep$_-$standardmodel.gif}
\bibitem{} 
  P. A. G. Scheuer,
  \textit{How high can a mountain be},
  JApA, 2, 165-169, 1981
\bibitem{} 
  P. B. Pal,
  {\textit At the root of things},
  Resonance, 14(6), 544-567, 2009
\end{thebibliography}
\end{document}